\documentclass[apjl]{emulateapj}
\usepackage{apjfonts}

\usepackage{graphicx}

\usepackage{amsmath}

\shorttitle{
3D MHD TURBULENCE BEHIND RELATIVISTIC SHOCK WAVES
}
\shortauthors{T. INOUE ET AL.}

\begin{document}

\title{
THREE-DIMENSIONAL SIMULATIONS OF MAGNETOHYDRODYNAMIC TURBULENCE BEHIND RELATIVISTIC SHOCK WAVES AND THEIR IMPLICATIONS FOR GRBs
}
\author{Tsuyoshi Inoue\altaffilmark{1,2}, Katsuaki Asano\altaffilmark{3}, and Kunihito Ioka\altaffilmark{4}}
\altaffiltext{1}{Department of Physics and Mathematics, Aoyama Gakuin University, Fuchinobe, Chuou-ku, Sagamihara 252-5258, Japan; inouety@phys.aoyama.ac.jp}
\altaffiltext{2}{Division of Theoretical Astronomy, National Astronomical Observatory of Japan, Osawa, Mitaka 181-8588 Japan}
\altaffiltext{3}{Interactive Research Center of Science, Graduate School of Science, Tokyo Institute of Technology, Ookayama, Meguro-ku, Tokyo 152-8550, Japan}
\altaffiltext{4}{KEK Theory Center and the Graduate University for Advanced Studies, Oho, Tsukuba 305-0801, Japan}

\begin{abstract}
Relativistic astrophysical phenomena such as gamma-ray bursts (GRBs) and active galactic nuclei often require long-lived strong magnetic field that cannot be achieved by shock compression alone.
Here, we report on three-dimensional special-relativistic magnetohydrodynamic (MHD) simulations that we performed using a second-order Godunov-type conservative code, to explore the amplification and decay of macroscopic turbulence dynamo excited by the so-called Richtmyer-Meshkov instability (RMI; a Rayleigh-Taylor type instability).
This instability is an inevitable outcome of interactions between shock and ambient density fluctuations.
We find that the magnetic energy grows exponentially in a few eddy-turnover times, because of field-line stretching, and then, following the decay of kinetic turbulence, decays with a temporal power-law exponent of $-0.7$.
The magnetic-energy fraction can reach $\epsilon_B \sim 0.1$ but depends on the initial magnetic field strength, which can diversify the observed phenomena.
We find that the magnetic energy grows by at least two orders of magnitude compared to the magnetic energy immediately behind the shock, provided the kinetic energy of turbulence injected by the RMI is larger than the magnetic energy.
This minimum degree of the amplification does not depend on the amplitude of the initial density fluctuations, while the growth timescale and the maximum magnetic energy depend on the degree of inhomogeneity in the density.
The transition from Kolmogorov cascade to MHD critical balance cascade occurs at $\sim 1/10$th the initial inhomogeneity scale, which limits the maximum synchrotron polarization to less than $\sim 2\%$.
We derive analytical formulas for these numerical results and apply them to GRBs.
New results include the avoidance of electron cooling with RMI turbulence, the turbulent photosphere model via RMI, and the shallow decay of the early afterglow from RMI.
We also performed a simulation of freely decaying turbulence with relativistic velocity dispersion.
We find that relativistic turbulence begins to decay much faster than one eddy-turnover time because of fast shock dissipation, which does not support the relativistic turbulence model by Narayan \& Kumar.
\end{abstract}


\section{INTRODUCTION}
It is well known that magnetic fields play a very important role in high-energy astrophysical phenomena such as particle acceleration and synchrotron emission.
Strong magnetic fields are often required around shock waves to explain the observed emissions.
For example, at the shocks in supernova remnants (SNRs), gamma-ray bursts (GRBs), and active galactic nuclei (AGNs), magnetic field strengths with orders of magnitude larger than their ambient fields are needed, and these cannot be achieved by only shock compression \citep[e.g.,][]{ML}.
Thus, some amplification mechanisms have been proposed such as the Weibel instability \citep{W59}, the cosmic-ray streaming instability \citep{LB, BL}, and the (small-scale) dynamo effect \citep{B50} due to the turbulence induced by magnetohydrodynamic (MHD) instabilities \citep{BBC, GJ, IYI09, IYI10, ZMW, MPNZNH}.
Each mechanism has its own advantage.
The Weibel instability can create magnetic fields even from unmagnetized media, and the cosmic-ray streaming instability can amplify upstream magnetic fields that are essential in first-order Fermi-acceleration processes in the shock wave.
Because the dynamo effect works under the influence of macroscopic hydrodynamic instabilities, it can generate larger scale magnetic field fluctuations than those induced by the microscopic plasma instabilities, which may be essentially important for synchrotron radiation and scattering high-energy particles.
Using three-dimensional relativistic MHD simulations, \cite{ZMW} showed that the Kelvin-Helmholtz instability (KHI) induced by relativistic shear flows activates turbulence dynamos.

In this study, we investigate the turbulent-dynamo effect with three-dimensional relativistic MHD simulations by taking GRBs as an example.
We study the effects of the Richtmyer-Meshkov instability (RMI), which are induced when the preshock density is inhomogeneous \citep[see][for reviews of RMI]{B02, NWMIZ}.
The RMI is a Rayleigh-Taylor-type instability that deforms the shock front and leaves vorticity behind the shock waves.
For the Rayleigh-Taylor instability, deformation of a discontinuity is triggered by gravitational acceleration, while the RMI is triggered by an impulsive acceleration caused by the passage of shock.

Because GRBs are believed to be associated with relativistic shocks of intermittent and inhomogeneous outflows \citep{mes06}, we can reasonably expect the work of RMI.
In the framework of Newtonian fluid dynamics, it was shown using MHD simulations in the context of SNRs that the preshock density inhomogeneity cause the magnetic field amplification \citep{GJ, IYI09, IYI10}.
In a relativistic shock wave, the idea was examined analytically by \cite{SG}, who considered how vorticity is generated because of the interaction between an ultra-relativistic shock and a density bump.
Recently, using two-dimensional special relativistic MHD simulations, \cite{MPNZNH} showed that the magnetic field can be amplified in the postshock of an inhomogeneous medium.
However, it is known that turbulence in two dimensions usually leads to very intermittent structures of velocity and magnetic fields compared to the three-dimensional case due to inverse cascade of enstrophy \citep{B08} that may induce ``fake'' magnetic field amplifications.
Thus, the evolution of turbulence and resulting magnetic field amplification should be verified by three-dimensional simulations, although the result of \cite{MPNZNH} is suggestive.

In this paper, using three-dimensional special relativistic simulations, we examine the following items:
(i) the generation of turbulence at the relativistic shock front through the RMI and the resulting magnetic field amplification caused by turbulence and 
(ii) the decay of relativistic MHD turbulence based on the results of (i).
Because it is computationally very expensive to follow the long-term evolution of turbulence in the former simulations, we cannot obtain the saturation level or decay rate of the turbulent magnetic field.
The supplementary simulations of (ii) allow us to study the decaying phase of turbulence as a consequence of its long-term evolution.

This paper is organized as follows.
In \S\ref{sec:setup} we provide numerical techniques and setups.
The results of the numerical simulations are presented in \S\ref{sec:RMI} (generation of turbulence and growth of magnetic field) and \S\ref{sec:decay} (long-term evolution and turbulence decay).
In \S\ref{sec:sum}, we summarize our findings from the simulations.
Finally,  in \S\ref{sec:GRB}, we discuss the implications of the GRB phenomenologies, including the internal shock synchrotron model, the photosphere model, the afterglow, and polarizations of these phenomena.

\section{NUMERICAL SETUP}\label{sec:setup}
We solve the ideal special-relativistic MHD equations.
Because we treat propagation of relativistic shock waves and turbulent magnetic fields, a high-resolution shock-capturing scheme as well as a divergence-free induction-equation solver are preferred.
We have developed such a code by using the five-wave Harten-Lax-van Leer Riemann solver \citep{MK05} developed by \cite{MUB09} and the constrained transport scheme \citep{EH} designed by \cite{SG09}.
The five-wave HLL solver employed in this paper is also called the HLLD solver.
According to \cite{BS11}, HLLD solver is more suited to treat magnetic field amplification by turbulence than other HLL solvers.
We impose the TM equation of state (EOS) proposed by \cite{MPB05}, which correctly describes the EOS for nonrelativistic and relativistic temperature limits and can describe the transition regime that differs from the exact EOS by less than 4\%.
We solve the equations in the conservative fashion so that the mass, momentum, and energy are conserved to within the round-off error.
Because we solve the scale-free, ideal (relativistic) MHD equations, we use a dimensionless time normalized by the light-crossing-time of the numerical domain $\tilde{t}=t\,c/L_z$, where $L_z$ is the length of the numerical box along the $z$-axis.
Thus, for example, if we choose the scale $L_{z}=10^7$ cm, $\tilde{t}=1$ corresponds to 0.33 msec.
For intuitive presentation, we treat variables such as density, velocity, pressure, and magnetic field dimensionally.
Note that the choice of $L_{z}$ does not change the value of the above variables--it only affects the timescale.

\begin{table}
\begin{center}
\caption{Model list\label{tbl1}}
\begin{tabular}{clllll}
\tableline\tableline
Run name& Density dispersion & Magnetic field strength$^a$ \\
\tableline
A1 & $\Delta n/n_0$ = 0.9 & 7.50 G \\
A2 & 0.6 & 7.50 G \\
A3 & 0.3 & 7.50 G \\
A4 & 0.1 & 7.50 G \\
\tableline\tableline
Run name& Velocity dispersion$^b$ & Magnetic field strength$^c$ \\
\tableline
B1 & $\Delta v/c_{\rm s}$ = 1.0 & 2170 G \\
B2 & 1.0 & 217 G \\
B3 & 1.0 & 21.7 G \\
B2-s & 0.3 & 217 G \\
B2-r & 1.7 ($\langle \Gamma \rangle=3.0$) & 217 G \\
\tableline
\tablenotetext{a}{Initial magnetic field strength in the fluid rest frame of the colliding flows.}
\tablenotetext{b}{$c_{\rm s}=0.48\,c$ is the speed of sound in the initial medium.}
\tablenotetext{c}{Initial magnetic field strength in the simulation frame, i.e., the postshock rest-frame when upstream is uniform.
Thus, the initial magnetic field strengths shown correspond to the values obtained after the compression by internal shock.}
\end{tabular}
\end{center}
\end{table}

\subsection{Initial Conditions of the First Set of Simulations}
In the first set of simulations, we prepare a cubic domain whose aspect ratio is $L_x:L_y:L_z=2:1:1$, which is divided by $N_{\rm x}\times N_{\rm y}\times N_{\rm z}=512\times 256 \times 256$ finite uniform volume elements.
We initially generate an inhomogeneous medium by adding isotropic sinusoidal density fluctuations in the fluid rest frame:
\begin{eqnarray}
n=n_0+\sum_{k_x,k_y,k_z}P(k)^{1/2}\,\sin(k_x\,x+k_y\,y+k_z\,z+\phi_k),
\end{eqnarray}
where $\phi_k$ is a random phase that depends on the wave number.
The mean number density of the medium is set to $\langle n \rangle =n_0=10^{10}$ protons cm$^{-3}$, which is compatible with GRB shells that collides at $r\sim 10^{14}$ cm from the central engine.
The fluctuations are characterized by their power spectrum $P(k)$.
We fix the spectral shape as a flat spectrum $k^2\,P(k)\propto k^0$ cutoff at the scale 
\begin{equation}
\lambda_{\rm c}\equiv \frac{2\pi}{k_{\rm c}}=\frac{L_{z}}{3},
\end{equation}
so that the fluctuations are only in large scales.
Note that the summations are performed for nonzero $k=(k_x^2+k_y^2+k_z^2)^{1/2}$.
We examine four situations that have density dispersions $\Delta n/\langle n \rangle=$0.1, 0.3, 0.6, and 0.9, where $\Delta n=\sqrt{\langle n^2\rangle-\langle n\rangle^2}$.

To induce shock waves, we set the converging velocity field to $v_{x}=0.818\,c$ at $x<L_x/2$ and $v_{x}=-0.818\,c$ at $x>L_x/2$, where $c$ is the speed of light, which indicates that the simulations are performed in a postshock rest frame when the upstream is uniform.
The Lorentz factor of the flows in the simulation frame is $\Gamma=1.74$ (the relative Lorentz factor of the flows is $\bar \Gamma=5.05$), which corresponds, for example, to the collision of the GRB shells with $\Gamma_{\rm lab}=1000$ and $100$ in the laboratory frame.

The pre-shock temperature of the GRB shell in the fireball model is usually too cold for our conservative numerical scheme to perform stable simulations (see eq. [\ref{eq:preT}]).
Thus, we increases the temperature to 9.38 MeV for the purpose of numerical stability.
By solving shock-jump conditions, we have confirmed that the discrepancy of the average postshock conditions (density, pressure, and magnetic field) between the case of this warm preshock temperature and the cold-limit case ($T=0$ K) is only 3\% because our initial temperature is still much lesser than the postshock temperature of $\sim 1$ GeV.
Thus, we stress that our initial condition is sufficiently cold to generate realistic GRB internal shock propagation.

The initial magnetic field strength in the fluid rest frame is fixed at $B=7.50$ G ($B=13.0$ G in the simulation frame) oriented in the $+y$ direction.
As discussed in \S6, if we assume the frozen-in transport of magnetic field from the central engine, this initial strength corresponds to a central engine magnetic field of $\sim 4\times10^{10}$ G (see, eq. [\ref{eq:B0}]).
Periodic boundary conditions are imposed for the $y$ and $z$ boundaries.
At the $x$ boundary, we continue to impose the converging relativistic flows in which density fluctuations same as the initial fluctuations are periodically input.
The initial model parameters and the names of the runs are summarized in Table 1 (see also \S\ref{sec:GRB} for the corresponding GRB initial conditions).

\subsection{Initial Conditions of the Second Set of Simulations}
As shown in the next section, the set of simulations discussed in the preceding section can only treat the generation of turbulence and the growing phase of the magnetic field, because the induced shock waves rapidly exit the computational domain.
To study the decay of turbulence and magnetic field, we need to perform long-term simulations.
To do so, we perform the following set of simulations:
We prepare a cubic domain with $L_x: L_y: L_z=1:1:1$ having resolution $N_{\rm x}\times N_{\rm y}\times N_{\rm z}=512^3$.
We fill the domain with an uniform medium whose density and temperature are $n=10^{11}$ cm$^{-3}$ and $k_{\rm B}\,T=0.235$ GeV, respectively, which are compatible with the averaged postshock values obtained from the first set of simulations.
Turbulence is initially given by summing the sinusoidal velocity fluctuations with zero mean velocity:
\begin{equation}
\Gamma\,v_i=\sum_{k_x,k_y,k_z}P(k)^{1/2}\,\sin(k_x\,x+k_y\,y+k_z\,z+\phi_{k,i}),
\end{equation}
where the subscript $i$ covers $x,\,y$, and $z$, and $\phi_{k,i}$ is a random phase that depends on the wave number and the components.
Note that the summation is overall nonzero $k=(k_x^2+k_y^2+k_z^2)^{1/2}$, and we generate fluctuating fields of $\Gamma\,v_i$ to avoid creating superluminal regions.
Their power spectrum is the flat with cutoff scale 
\begin{equation}
\lambda_{\rm c}\equiv \frac{2\pi}{k_{\rm c}}=\frac{L_{z}}{4}.
\end{equation}
In the generated initial velocity field thus generated, the ratio of the total power of the rotational velocity component ($\vec{\nabla}\cdot\vec{v}=0$) to the compressive component ($\vec{\nabla}\times\vec{v}=0$) is approximately 2:1, which indicates that the initial turbulence is dominated by incompressible flows.
The periodic boundary conditions are imposed and we perform several runs with different initial velocity dispersions ($\Delta v\equiv \sqrt{\langle v^2 \rangle}$) and initial magnetic field strengths, whose values are summarized in Table 1 (see also \S\ref{sec:GRB} for the corresponding GRB initial conditions).

\section{SHOCK PROPAGATION IN INHOMOGENEOUS MEDIA}\label{sec:RMI}
\subsection{Generation of Turbulence}
The initial converging flows used in the series of run A induce two oppositely propagating shock waves.
When the shock wave passes a density bump or dent, a RMI is induced and leaves vorticity behind the shock waves.
For the Rayleigh-Taylor instability, deformation of a discontinuity and generation of vorticity are triggered by gravitational acceleration, while the RMI is triggered by an impulsive acceleration caused by the passage of the shock wave \citep{B02, NWMIZ}.

\begin{figure}[t]
\epsscale{1.35}
\plotone{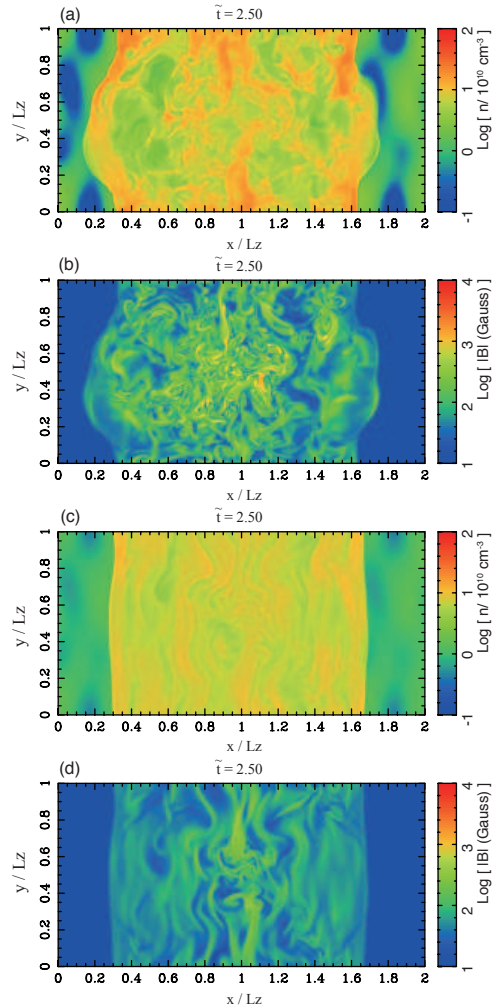}
\caption{
(a): Two-dimensional number density slice of the result of run A1.
(b): Two-dimensional magnetic field strength slice of the result of run A1.
(c): Two-dimensional number density slice of the result of run A3.
(d): Two-dimensional magnetic field strength slice of the result of run A3.
All panels show the structures at $z=0.0$ and $\tilde{t}=2.5$.
Physical quantities are measured in the simulation frame, which is equivalent to the postshock rest frame when the upstream is uniform.
}
\label{f1}
\end{figure}

\begin{figure}[t]
\vspace{-13cm}
\epsscale{2.0}
\plotone{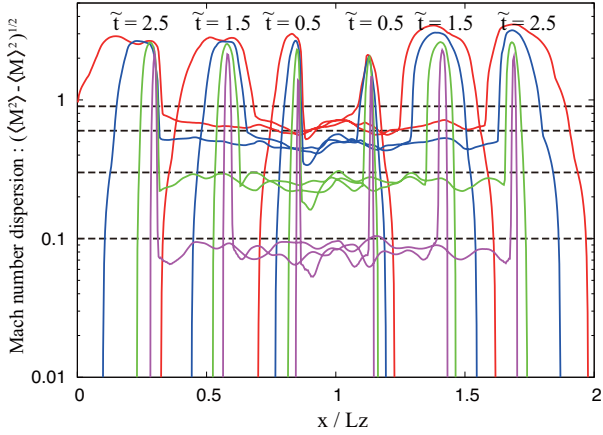}
\caption{
Dispersion of the sonic Mach number in the $y$-$z$ plane at a given $x$.
The red, blue, green, and magenta lines represent the results of runs A1, A2, A3, and A4, respectively.
Series of the same-color lines show different temporal snapshots at $\tilde{t}=0.5,\,1.5,\,$ and $2.5$.
Also plotted as dashed lines are the reference lines for $M=0.1,\,0.3,\,0.6,$ and 0.9.
}
\label{f2}
\end{figure}

Fig.~\ref{f1} (a) and (c) show two-dimensional number-density slices of the results of runs A1 and A3, in which one can see the deformed shock fronts and turbulent density fields due to the RMI.
In the following, we plot the values measured in the simulation frame in figures, which is equivalent to the postshock rest frame when the upstream is uniform.
Since lower (higher) upstream density leads to faster (slower) shock propagation, the medium with larger density dispersion induces larger shock deformation, and thus leaves stronger vorticity or turbulence.
To clarify this, we plot the dispersion of the sonic Mach number in the $y$-$z$ plane at a given $x$ in Fig.~\ref{f2}.
The red, blue, green, and magenta lines show the results of runs A1, A2, A3, and A4, respectively.
Series of the same color lines represent different temporal snapshots at $\tilde{t}=0.5,\,1.5,\,$ and $2.5$.
We also plot the reference lines of $M=0.1,\,0.3,\,0.6,$ and 0.9 as dashed lines.
It is obvious that larger preshock density dispersion introduces stronger postshock velocity dispersion, and that the velocity dispersion is roughly proportional to the preshock density dispersion.
As the postshock Mach number dispersion approaches unity, in particular run A1, the Mach number dispersion becomes somewhat smaller than that expected from the linear dependence on the density dispersion.
This would be due to the formation of ``eddy shocklets" in the turbulent postshock \citep{KO90} that tend to keep the velocity dispersion subsonic.

\cite{SG} developed an analytic formula for the generation of vorticity because of an interaction between a relativistic shock and a density bump using the geometric shock-dynamics approximation \citep[see, also][]{GM}.
They found that the vortical energy density generated by the interaction is proportional to the square of the maximum density of the bump (in other words, the velocity dispersion of vorticity is proportional to the density contrast), when the density contrast is small.
Because they assumed ultra relativistic shock wave to study the afterglow phase of GRBs, it may be inappropriate to compare their result and that of the mildly relativistic case examined here.
However, note that our result (linear dependence of the velocity dispersion on the preshock density dispersion when $\Delta n/n_0\ll 1$) is consistent with their finding.
In addition to the density bump-shock interaction, also note that the interaction with density dents also generates vorticity because the RMI is known to be effective even in that case. 

Fig.~\ref{f2} shows that even when $\tilde{t}=2.5$ the level of the post shock velocity dispersion at around $x/L_z=1.0$ is almost the same as for $\tilde{t}=0.5$, which indicates that the turbulence has not yet begun to decay.
In the present simulations, the postshock velocity dispersion is nonrelativistic, and in Newtonian fluid dynamics, it is widely known that the turbulence without a continuous driving source begins to decay within a few eddy-turnover times.
Because the average postshock sound speed is $c_{\rm s}\simeq 0.5\,c$ and the scale of vortices that is given essentially by the scale of the density fluctuations is $\sim L_{z}$, the eddy-turnover time can be estimated to be a few light-crossing times of $L_z$.
Thus, to confirm the decay of turbulence, we need to continue the simulation for several light-crossing times, which requires a larger numerical domain.
The decaying phase of turbulence will be shown in \S4 as the results of the second set of simulations.

\subsection{Magnetic Field Amplification}
In addition to the shock compression, the turbulence induced by the RMI amplifies the magnetic field by stretching the field lines.
The slices of magnetic field strength distribution of runs A1 and A3 are shown in Fig.~\ref{f1} (b) and (d), respectively.
We can confirm the magnetic field amplification far beyond the value achieved by the shock compression alone (if the preshock medium is uniform, the magnetic field strength is $B\sim 70$ G).

\begin{figure}[t]
\vspace{-6.5cm}
\epsscale{1.8}
\plotone{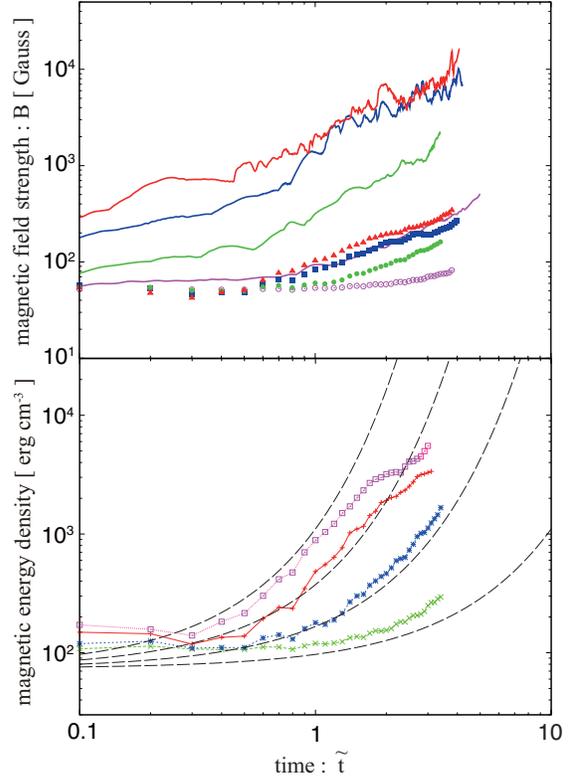}
\caption{
Top: Evolution of the maximum ($|B|_{\rm max}$: lines) and average ($\langle |B| \rangle$: points) magnetic field strengths.
Bottom:  Evolution of average magnetic energy density at the $x/L_{z}=1.0$ plane ($\langle \Gamma^2\,B^2/8\pi \rangle_{x=1}$).
Red, blue, green, and magenta lines show the results of runs A1, A2, A3, and A4, respectively.
Dashed lines show the model evolution eq. (\ref{Bevo}) with $f=1.8$.
}
\label{f3}
\end{figure}

Similar amplification of magnetic fields in the dynamics of SNRs formed in various ambient ISM has been reported by many authors (see Balsara et al. 2001 for turbulent ISM; Giacalone \& Jokipii 2007 for ISM whose density fluctuation power spectrum is a power-law, which is recently applied to relativistic shocks by Mizuno et al. 2010; Inoue et al. 2009, 2010 for cloudy ISM).
The initial density fluctuations for our simulations are similar to that of \cite{GJ} and \cite{MPNZNH}.
However, they examined shock propagation in a two-dimensional geometry, and it is widely known that a two-dimensional turbulence can be qualitatively different from a three-dimensional turbulence because of the inverse cascade of enstrophy, which creates long-lived, large-scale eddies.
At later stages, the large-scale eddies decay very intermittently and leads to a ``fake" magnetic field amplification \citep[see, e.g.,][]{B08}.
Thus, the amplification reported in the previous two-dimensional simulations seems suspicious \citep{BJL}.
Nevertheless, such a dimensional difference stands out in the later-stage of driven turbulence, whereas the amplifications caused by density fluctuation-shock interaction is a short-term phenomenon.
The presence of the magnetic field amplifications in our three-dimensional simulations suggests that the amplifications reported in the previous two-dimensional simulations are not fake, although the results of the two-dimensional simulations may be affected by the intemittency in some degree.
Recently, \cite{IYI09, IYI10} also showed that similar magnetic field amplification caused by shock-cloud interactions does not depend on spatial dimensions.

We plot the evolution of the maximum and average magnetic field strengths ($|B|_{\rm max}$ and $\langle |B| \rangle$) in the top panel of Fig.~\ref{f3} and that of average magnetic energy density at the $x/L_{z}=1.0$ plane ($\langle \Gamma^2\,B^2/8\pi \rangle_{x=1}$) in the bottom panel, where $\Gamma$ is the Lorentz factor of each fluid element.
Note that, since the post shock turbulence is subsonic and thus nonrelativistic, $\Gamma$ in the above definition of the magnetic energy density is approximately unity.
It is clear that the amplification still continues, since the turbulence that is the source of the amplification has not yet decayed.
Since the magnetic field is passive with respect to the turbulent velocity field, we can reasonably expect from the induction equation [$\partial \vec{B}/\partial t=\vec{\nabla}\times(\vec{v}\times\vec{B})\simeq \Delta v\,\vec{B}/l_{\rm turb}$] that the magnetic energy density evolves as
\begin{equation}
\langle e_B \rangle \simeq \langle e_B\rangle_{\rm ini}\,\exp\left(\frac{f\,\Delta v}{l_{\rm turb}}\,t\right),\label{Bevo}
\end{equation}
where $\langle e_B \rangle_{\rm ini}=\langle \Gamma^2\,B^2 \rangle_{\rm ini}$ is the initial magnetic energy density (immediately behind the shock), $l_{\rm turb}$ is the scale of the turbulent flow, $\Delta v$ is the velocity dispersion of the turbulence, and $f$ is a factor on the order of unity.
In the present simulations, the turbulent flows with the scale $l_{\rm turb}=L_z/3$ would give the fastest growing mode, because the scale of the flows is essentially determined by the scale of the initial density fluctuations.
From Fig.~\ref{f2}, the velocity dispersion can be evaluated as $\Delta v/c_{\rm s}\sim 0.9,\,0.6,\,0.3,$ and $0.1$ for the runs A1, A2, A3, and A4, respectively.
In the bottom panel of Fig.~\ref{f3}, we plot the evolution of this simple model with $f=1.8$ as dashed lines.
We see that the model equation reproduces the growth rate especially for $\tilde{t}\sim1$.
In the next section, we show that the same model with $f=1.8$ also reproduces the initial growth phase of magnetic energy, even for the second set of simulations.

\section{DECAY OF TURBULENCE AND MAGNETIC FIELD}\label{sec:decay}
In the previous section, we have seen that the density fluctuation-shock interactions generate turbulence whose velocity dispersion can be as large as post shock sound speed depending on the amplitude of the density fluctuations.
In this section, we show the results of the second set of simulation.
The initial conditions of the run B1, B2, B3, and B2-s, which have transonic and subsonic velocity dispersions, simulate the post shock turbulence induced by the density fluctuation-shock interactions.
We can study the long-term evolution of turbulence and magnetic field from these simulations.

The run B2-r has a relativistic initial velocity dispersion ($\langle \Gamma \rangle=3.0$) that may not be created by the density fluctuation-shock interactions.
However, since there were several discussions on the possibility of relativistic turbulence in GRB \citep{L06,nar09,KN,L09}, it would be meaningful to study it.

\subsection{Transonic and Subsonic Turbulence}
\subsubsection{Evolution Process}\label{sec:evolve}

\begin{figure}[t]
\vspace{-4.cm}
\epsscale{1.2}
\plotone{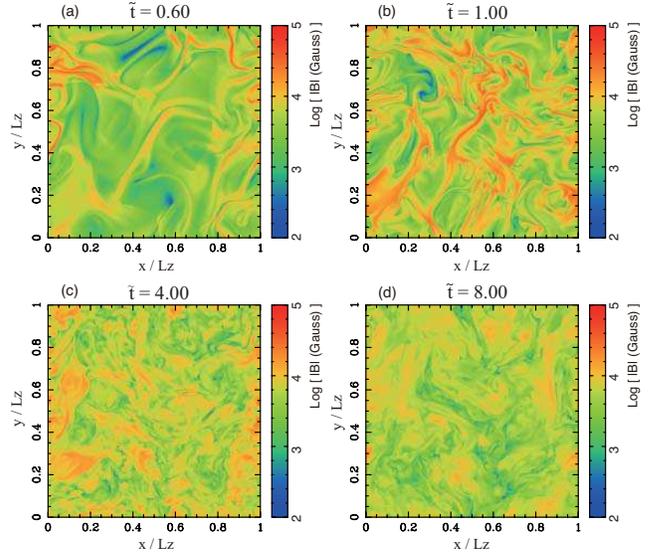}
\caption{
Slices of magnetic field strength of run B1 in the $z=0.0$ plane.
Panels (a), (b), (c), and (d) represent temporal snapshots at $\tilde{t}=0.6,\,1.0,\,4.0$, and $8.0$, respectively.
}
\label{f4}
\end{figure}

\begin{figure}[t]
\vspace{-7.5cm}
\hspace{1.0cm}
\epsscale{2.0}
\plotone{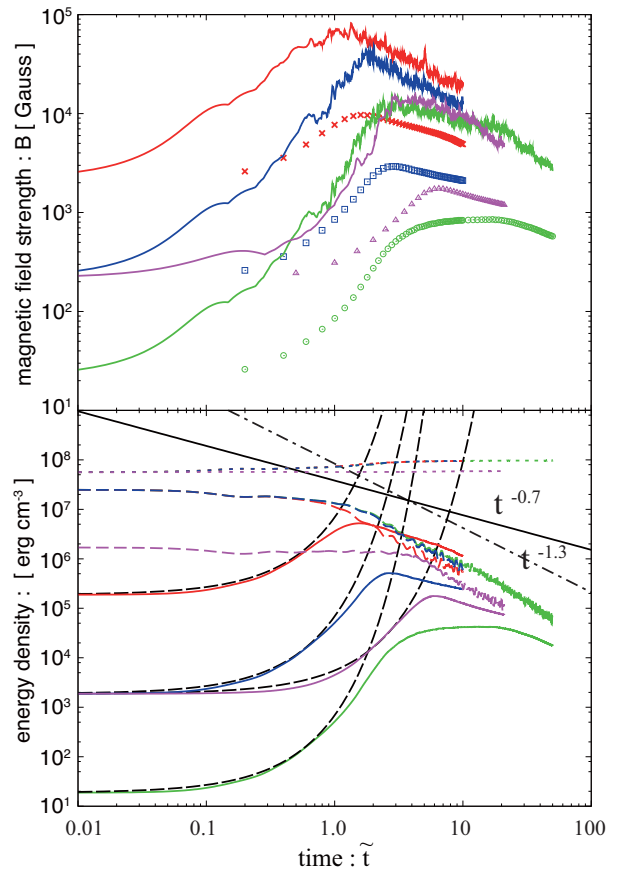}
\caption{
Top panel: evolution of the maximum and the average magnetic field strengths ($|B|_{\rm max}$ and $\langle |B| \rangle$) obtained from runs B1(red), B2 (blue), B3 (green), and B2-s (magenta).
Solid lines and points respectively indicate the maximum and average field strengths, respectively.
Bottom panel: evolution of magnetic (solid), kinetic (dashed) and internal (dotted) energy densities.
Line colors identify the models as per the top panel.
Dashed black lines show the model of the evolution of the magnetic energy density given by eq. (\ref{Bevo}) with $f=1.8$.
The thin black line is a reference line proportional to $t^{-0.7}$ fit to the decay-phase evolution of the magnetic energy, and the dot-dashed black line is a reference line proportional to $t^{-1.3}$ fit to the decay-phase evolution of the kinetic energy.
}
\label{f5}
\end{figure}

As in the first set of simulations, the magnetic field in the turbulence is amplified by the field-line stretching.
In Fig.~\ref{f4}, we plot the $z=0$ plane slices of the magnetic field strength for run B1.
Panels (a), (b), (c), and (d) represent temporal snapshots at $\tilde{t}=0.6,\,1.0,\,4.0$, and $8.0$, respectively.
The evolution of the maximum and the average magnetic field strengths ($|B|_{\rm max}$ and $\langle |B| \rangle$) obtained from runs B1(red), B2 (blue), B3 (green), and B2-s (magenta) is plotted in the top panel of Fig.~\ref{f5}.
Solid lines and points indicate the maximum and average field strengths, respectively.
In the bottom panel, we also plot the magnetic (solid), kinetic (dashed) and internal (dotted) energy densities, where we have defined them $\langle \Gamma^2\,B^2/8\pi \rangle$, $\langle \rho\,c^2\,\Gamma\,(\Gamma-1) \rangle$, and $\langle \{\Gamma^2\,\gamma/(\gamma-1)-1\}\,p \rangle$, respectively.
Note that their summation is conserved to within the round-off error owing to the conservative numerical scheme.
For runs B1, B2, and B3, the kinetic energy begins to decay at $\tilde{t}\sim2$ (i.e., a few eddy-turnover times), and the magnetic energy also begins to decay when it becomes comparable to the decaying kinetic energy.
The initial velocity dispersion of run B2-s is approximately one-third of runs B1, B2, and B3, which results in a larger eddy-turnover time and thus kinetic energy decay postponed by a factor of approximately three.
Because magnetic energy does not dominate energy budget, results with different initial magnetic field strengths (i.e., runs B1, B2, and B3) show almost the same evolution for the kinetic and internal energies.

\begin{figure}[t]
\vspace{-11.5cm}
\epsscale{1.8}
\plotone{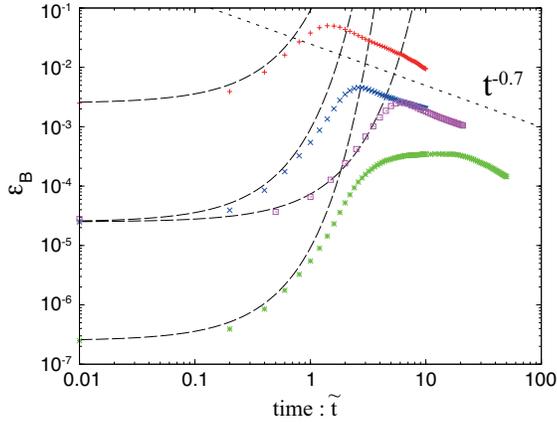}
\caption{Evolution of $\epsilon_B$ obtained from runs B1(red), B2 (blue), B3 (green), and B2-s (magenta).}
\label{f6}
\end{figure}

Until the turbulence begins to decay, eq. (\ref{Bevo}) with $f=1.8$ (where we have substituted the initial velocity dispersion as $\Delta v$ and $l_{\rm turb}=L_z/4$ from the initial condition), which is plotted in Fig.~\ref{f5} with a thin dashed line, again gives a suitable model.
Note that in the case of a turbulence dynamo with a continuous large-scale driving source, the linear growth stage of the magnetic energy is followed by the initial exponential growth stage that continues until the magnetic energy becomes comparable to the kinetic energy \citep{SC07, CVBLR09}.
In our simulations, the turbulence is injected only initially and decays eventually, which halts the magnetic field amplification and leads to damping.
In the decay phase, the evolution of the magnetic energy can be fit by a power law with an index $\sim -0.7$ and the kinetic energy also shows power-law decay with an index $\sim -1.3$.
The ratio of magnetic energy to internal energy, which is often denoted as $\epsilon_B$ is an important parameter in GRB-emission models.
We plot the $\epsilon_B$ for runs B1, B2, B3, and B2-s in Fig.~\ref{f6}.
Because the initial internal energy is comparable to or larger than the initial kinetic energy, the internal energy increases only slightly, even after the decay of turbulence.
Thus, the evolution of the $\epsilon_B$ mainly determined by the evolution of the magnetic energy, and the simple growth model obtained by dividing eq. (\ref{Bevo}) by the initial internal energy well fit the early-phase evolution (see, dashed lines in Fig.~\ref{f6}).
The later power-law decay with the exponent $-0.7$ (dotted line in Fig.~\ref{f6}) also fits the later evolution of $\epsilon_B$.

\begin{figure}[t]
\vspace{-7.3cm}
\epsscale{1.7}
\plotone{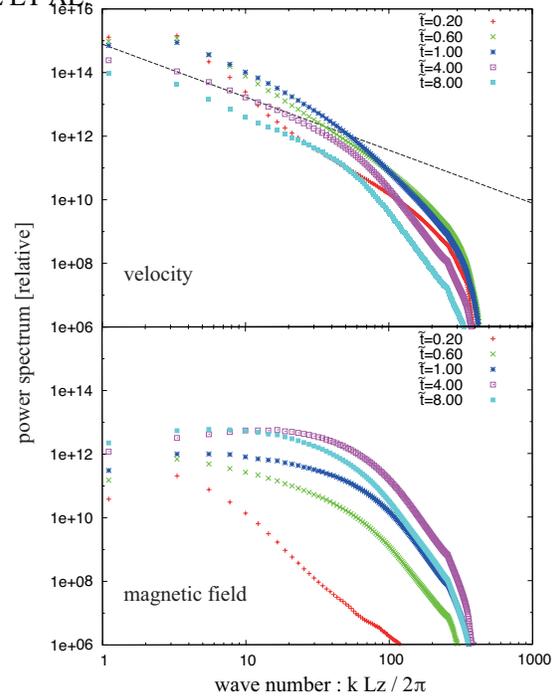}
\caption{
Power spectra of velocity $P_v(k)$ ({\it top}) and magnetic field $P_B(k)$ ({\it bottom}) at $\tilde{t}=0.2,\,0.6,\,1.0,\,4.0,$ and $8.0$ from run B2.
Here, the power spectra are defined as $\int P_v(k)\,dk=\int v^2\,d^3x$ and $\int P_B(k)\,dk=\int B^2\,d^3x$.
Dashed line shows the Kolmogorov spectrum ($P_v(k)\propto k^{-5/3}$).
}
\label{f7}
\end{figure}

Can we predict the maximum $\epsilon_B$ using parameters such as $\Delta v$ and $\epsilon_{B,{\rm ini}}$?
In the present simulations, we can describe the eddy-turnover time for the smallest initial eddy as $t_{\rm eddy}\simeq L_{z}/4\,\Delta v$ ($\simeq 0.5\,L_z/c$ for runs B1, B2, and B3, and $\simeq 1.7\,L_z/c$ for run B2-s, where we have used $c_{\rm s}\simeq 0.5\,c$).
Thus, as seen from the bottom panel of Fig. \ref{f5}, the turbulence can maintain its the initial strength until $t\simeq3\,t_{\rm eddy}$.
Substituting the above timescale into eq. (\ref{Bevo}), we obtain
\begin{equation}\label{Bamp}
\frac{\langle e_B(t_{\rm dec}) \rangle}{\langle e_{B}\rangle_{\rm ini}}\simeq\exp(\,3\,f\,)\sim 10^2.
\end{equation}
Note that this degree of amplification describes the ratio of the magnetic energy at $t=0$ (immediately behind the shock wave) and at the time when the turbulence begins to decay ($t_{\rm dec}$).
This degree of amplification is independent of the initial velocity dispersion that would explain the comparable maximum $\epsilon_B$ of run B2 and B2-s.
However, this degree of amplification cannot be used for the maximum $\epsilon_B$ prediction because the $\epsilon_B$ can grow, even after the $t_{\rm dec}$, until the magnetic energy becomes comparable to the kinetic energy, although after the $t_{\rm dec}$ the growth rate becomes slower than the exponential growth given by eq. (\ref{Bevo}).
Indeed, for run B3, the maximum $\epsilon_B$ is approximately three orders of magnitude larger than the initial value.
An accurate description of the growth of $\epsilon_B$ for $t>t_{\rm dec}$ may not be obtained in a straightforward way; nevertheless, we can predict the upper limit of the maximum $\epsilon_B$ as follows:
Because the growth of $\epsilon_B$ stops when the magnetic energy becomes comparable to the kinetic energy, the upper-limit of $\epsilon_B$ should be smaller than the ratio of the kinetic energy to the internal energy $\epsilon_{\rm K}$.
For $t>t_{\rm dec}$, the kinetic energy $e_{\rm K}$ evolves as $\propto t^{-1.3}$, which leads to the inequality,
\begin{equation}\label{BK}
\epsilon_{B, {\rm max}}\lesssim \epsilon_{\rm K}\simeq \langle M\rangle_{\rm ini}\,(1+t/t_{\rm dec})^{-1.3},
\end{equation}
where $ \langle M\rangle_{\rm ini}$ is the Mach number of the initial turbulence (see Fig. \ref{f5}).

\subsubsection{Structure of Magnetic Field}\label{Bstructure}

\begin{figure}[t]
\vspace{-12.7cm}
\epsscale{2.0}
\plotone{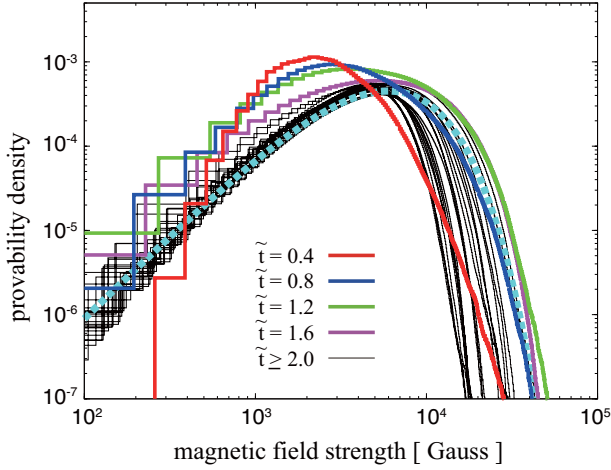}
\caption{
Probability distribution histograms of the run B1 at $\tilde{t}=0.4$ (red), $0.8$ (blue), $1.2$ (green), $1.6$ (magenta), and $\tilde{t}\ge2.0$ with the interval $\Delta\tilde{t}=0.4$ (thin black lines).
We also plot a distribution $\propto B^{2}\,\exp(-B/3\times 10^3)$ (light-blue dashed line) as a reference.
}
\label{f8}
\end{figure}

\begin{figure}[t]
\vspace{-20.cm}
\epsscale{2.8}
\plotone{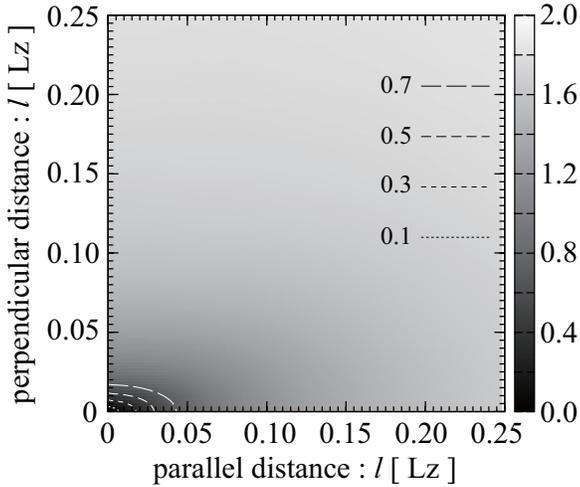}
\caption{
Structure function defined by $\langle |\vec{b}(\vec{r})-\vec{b}(\vec{r}+\vec{l})|^2\rangle_{\vec{r}}$ calculated by using the result of the run B1 at $\tilde{t}=4$, where $\vec{b}=\vec{B}/|\vec{B}|$.
The vertical axis $l_{\parallel}$ indicates the distance parallel to the local magnetic field from $\vec{r}$, and the horizontal axis $l_{\bot}$ indicates the perpendicular distance.
}
\label{f9}
\end{figure}

The spectra of velocity and magnetic field also evolve with time.
In Fig.~\ref{f7}, we plot the power spectra of the velocity $P_v(k)$ ({\it top}) and the magnetic field $P_B(k)$ ({\it bottom}) for $\tilde{t}=0.2,\,0.6,\,1.0,\,4.0,$ and $8.0$ of run B2, where the power spectra are defined as $\int P_v(k)\,dk=\int v^2\,dx^3$ and $\int P_B(k)\,dk=\int B^2\,dx^3$.
The amplitude of the spectra evolves along with the kinetic and magnetic energy densities, whereas their shapes are roughly maintained after $\tilde{t}\sim 1$ (after a few eddy-turnover times).
The shapes of spectra at $\tilde{t}\gtrsim 1$ resemble the steady spectra of super-Alfv\'enic turbulence with a large-scale continuous driving source \citep[e.g.,][]{CV00, CVBLR09, ZMW}.
For hydrodynamics, the spectrum of the rotational velocity component of the developed, decaying turbulence decreases its power while maintaining the Kolmogorov spectrum \citep{KO92}.
Recall that, in the series of run B, the initial power of the velocity field is dominated by the rotational component.
Thus, it is not surprising for the present spectra to evolve while maintaining their shapes.
The spectra of other models of run B also evolve in a qualitatively similar way.

In large scales ($k/2\pi\lesssim 30/L_z$), the velocity power spectrum for $\tilde{t}\gtrsim1$ exhibits a power law whose exponent is consistent with the Kolmogorov index of $-5/3$.
However, the velocity power spectrum becomes steeper roughly at the scale $\sim L_z/30$, even though it is much larger than the scale of numerical resolution $\simeq L_z/500$.
This would indicate the transition of the kinetic energy transfer mechanism from the hydrodynamic Kolmogorov cascade to the magnetohydrodynamic critical balance cascade \citep{GS95}.
The power spectrum of the magnetic field also changes its slope at the transition scale.
For large scales ($k/2\pi\lesssim 30/L_z$) $P_B$ for $\tilde{t}\gtrsim 1$ is roughly flat as commonly seen in the results of turbulence dynamo simulations \citep[][and references therein]{CV00, CVBLR09, ZMW, BS05}, and at small scales it becomes steeper.
The steepening of $P_B$ below the transition scale indicates that the magnetic energy is comparable to the kinetic energy at that scale.
Appearance of the transition scale is quite reasonable, because the velocity dispersion of turbulent eddies decreases as the eddies cascade to smaller scales and eventually vanishes, whereas magnetic field strength is not (i.e., there always exist a transition scale at which the velocity dispersion of eddies is comparable to the Alfv\'en velocity constructed using the strength of the mean magnetic field).
Thus, even if velocity dispersion is super-Alfv\'enic at large scales, it becomes sub-Alfv\'enic below the transition scale.

\begin{figure}[t]
\vspace{-6.cm}
\epsscale{1.6}
\plotone{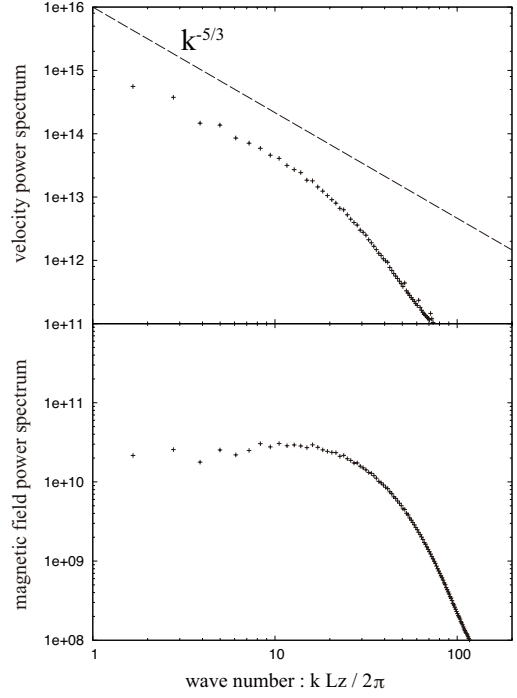}
\caption{
Power spectra of velocity $P_v(k)$ ({\it top}) and magnetic field $P_B(k)$ ({\it bottom}) of run A1 at $\tilde{t}=2.5$.
Dashed line shows the Kolmogorov spectrum.
}
\label{f10}
\end{figure}

The structure in the magnetic field strength in Fig. \ref{f4} suggests that the field strength distribution has large dispersion.
In Fig. \ref{f8}, we plot the probability distribution histograms of the magnetic field strength calculated using the result of run B1 at $\tilde{t}=0.4$ (red), $0.8$ (blue), $1.2$ (green), $1.6$ (magenta), and $\tilde{t}\ge2.0$ with the interval $\Delta\tilde{t}=0.4$ (thin black lines).
For $\tilde{t}\gtrsim1$, the distribution function can be fitted by the following function: $f(B)\propto B^{2}\,\exp(-B/a(t))$, where $a(t)$ is the function of time.
By taking the second moment of the normalized distribution $f(B)$ to be proportional to the decaying-phase magnetic energy density ($\propto t^{-0.7}$), we obtain $a(t)\propto t^{-0.35}$, which indicates that both the width and the peak of distribution decreases with time for $\tilde{t}\gtrsim1$ as expected.
We confirmed that the magnetic field strength distributions of the results of the other simulations (series of runs A and B) exhibit distributions similar to $f(B)$ given above.

Because the magnetic field is passive with respect to large-scale turbulent flows, the angular distribution of the magnetic field is nearly isotropic, whereas it shows a only slight dependence on the initial direction.
However, the isotropic distribution does not indicate the absence of local correlations of the magnetic field direction.
In Fig. \ref{f9}, we show the structure function defined by $\langle |\vec{b}(\vec{r})-\vec{b}(\vec{r}+\vec{l})|^2\rangle_{\vec{r}}$ calculated by using the result of run B1 for $\tilde{t}=4$, where $\vec{b}=\vec{B}/|\vec{B}|$.
The vertical axis $l_{\parallel}$ gives the distance from $\vec{r}$ parallel to the local magnetic field, and the horizontal axis $l_{\bot}$ gives the distance perpendicular to the local magnetic field.
If magnetic field orientation has spatial correlations, the structure function is null.
Fig. \ref{f9} shows that the magnetic field orientations have correlations on the small scale because the magnetic field is not passive below the transition scale.
The anisotropic structure along the local magnetic field direction is qualitatively consistent with the theory of anisotropic cascade of Alfv\'en waves \citep{GS95} and the result of turbulence dynamo simulations by \cite{CV00}.

Finally, we note here that the spectra from run A1 at $\tilde{t} =2.5$ that are plotted in Fig.~\ref{f10} are similar to those of run B2 at $\tilde{t}\gtrsim1$ in Fig.~\ref{f7}.
This similarity suggests that the second set of simulations (i.e., the series of run B) is appropriate for studying the long-term evolution of the turbulence driven by the RMI.

\subsection{Relativistic Turbulence}\label{sec:rela}

\begin{figure}[t]
\vspace{-7.cm}
\epsscale{2.}
\plotone{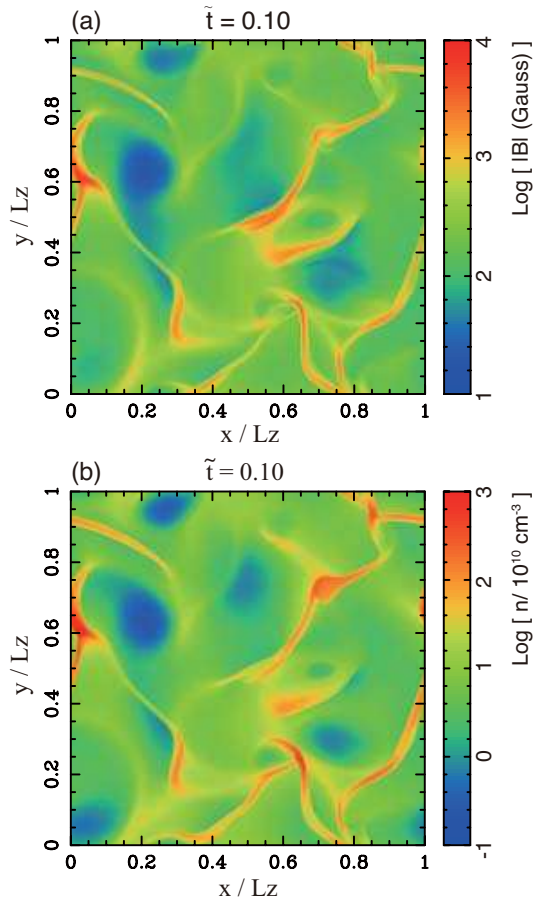}
\caption{
Two-dimensional slices at $z=0$ of magnetic field strength structure ({\it top}) and number density structure ({\it bottom}) from run B2-r at $\tilde{t}=0.10$.
Physical quantities are measured in the simulation frame.
}
\label{f11}
\end{figure}

\begin{figure}[t]
\epsscale{0.9}
\plotone{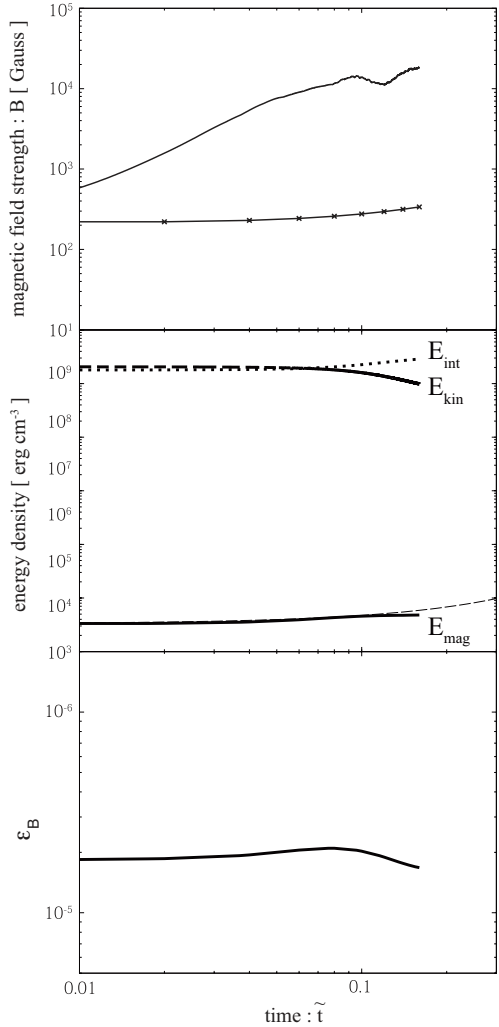}
\caption{
Top panel: Evolution of the maximum (solid) and average (line and points) magnetic field strengths.
Middle: Evolution of the magnetic (thick solid), kinetic (thick dashed) and internal (thick dotted) energies.
Thin solid line represents eq (\ref{Bevo}) with $f=1.8$.
Botom panel: Evolution of $\epsilon_B$.
}
\label{f12}
\end{figure}

For run B2-r, which has a relativistic initial velocity dispersion ($\langle\Gamma\rangle=3.0$), the magnetic field also grows with time.
We plot the $z=0$ plane slices of the magnetic field strength and number density for run B2-r at $\tilde{t}=0.10$ in Fig.~\ref{f11}.
In Fig.~\ref{f12}, we show the evolution of the maximum and average magnetic field strengths ({\it top}), the kinetic, magnetic, and internal energy densities ({\it middle}), and $\epsilon_B$ ({\it bottom}).
In run B2-r, the simulation is terminates at $\tilde{t}=0.16$ (slightly after the kinetic energy begins to decrease).
The reason for this is as follows:
Initially, we set the relativistic turbulent field in which the regions with oppositely oriented relativistic flows that tend to create vacua there always exist.
Because the employed numerical scheme is a conservative grid-based method, it is very difficult to treat a thin medium.
The continuation of the simulation might be possible, if we artificially put the mass and internal energy into the thin regions.
However, this strategy may significantly changes the dynamics because the volume-filling factor of such thin regions can be large.
Thus, we do not choose the continuation.

Although the timescale we have followed is short, the kinetic energy has already started to decay, whereas the magnetic energy is still in the growth phase.
The model equation (\ref{Bevo}) with $f=1.8$, $\Delta v=c$, and $l_{\rm turb}=L_{z}/4$, which is plotted as a dashed line in the middle panel of Fig.~\ref{f12}, again shows good agreement until the onset of the kinetic energy decay.
In the present case, the decay of the kinetic energy begins approximately ten times faster than that for the transonic cases, which we attribute to the shock dissipation that immediately converts kinetic energy into internal energy.
Indeed, a number of discontinuous density structures shown in Fig.~\ref{f11} indicate the formation of multiple shock waves.
The beginning of the kinetic energy decay much faster than the eddy-turnover time is consistent with the simulations of decaying supersonic turbulence \citep[e.g.,][]{MKBS}, although these are nonrelativistic, isothermal simulations.
Note that, when the kinetic energy injected into the turbulence is fixed, the timescale of the decay of the relativistic turbulence would substantially shorter than the nonrelativistic one, since the relativistic effect reduces the necessary mass for deceleration by $\Gamma^2$.

The evolution of $\epsilon_B$ is completely different from the transonic and subsonic cases (i.e., $\epsilon_B$ decreases with time) because the shock compression increases the internal energy more rapidly than the magnetic energy.
As for transonic and subsonic turbulence, the magnetic energy would be able to grow until it becomes comparable to the kinetic energy.
Thus, $\epsilon_B$ can be turn to increase once the velocity dispersion of turbulence becomes subsonic and the increase in the internal energy due to the shock dissipation ceases, provided the magnetic energy is smaller than the kinetic energy up to that time.

Therefore, the relativistic turbulence model for GRBs \citep{L06,nar09,KN,L09} does not seem to work because the relativistic turbulence decays much more rapidly than the eddy-turnover time $t_{\rm dec}\ll t_{\rm eddy}\sim L_z/c$ \citep[see also][]{ZJM}.

\section{SUMMARY OF SIMULATIONS AND COMPARISONS WITH PREVIOUS STUDIES}\label{sec:sum}
Using three-dimensional special-relativistic MHD simulations, we have studied the generation of turbulence as a consequence of the interactions between a relativistic shock  wave and density fluctuations.
We have found that the magnetic field is amplified due to the turbulence dynamo effect and follows a power-law decay in later stages.
The following items are noteworthy:
\begin{itemize}
\item The velocity dispersion of turbulence induced by the density fluctuation-shock interaction roughly depends linearly on the dispersion of the preshock density inhomogeneity.
It can be as large as the postshock sound speed when the dispersion of density fluctuations is on the order of the mean value ($\Delta n/n\sim1$).
Therefore, even if the shock is relativistic, the induced turbulence in our situation cannot be as highly relativistic as postulated in \citet{nar09} model.
\item For transonic and subsonic turbulence, the induced turbulence maintains its strength for a few eddy-turnover times and then decays.
In such a turbulent medium, until the turbulence begins to decay, the magnetic field is amplified exponentially in time according to eq. (\ref{Bevo}) because of the effect of field-line stretching.
The degree of amplification of the magnetic energy before the start of the turbulence decay is approximately 100, regardless of the initial velocity dispersion of turbulence (eq. [\ref{Bamp}]). 
The magnetic field continues to grow until the magnetic energy becomes comparable to the kinetic energy (eq. [\ref{BK}]) after which the magnetic energy follows a power-law decay with an exponent of $\sim 0.7$ (i.e., $e_B\propto t^{-0.7}$).
\item The evolution of $\epsilon_B$ (the ratio of the magnetic energy density to the internal energy density) in the transonic and subsonic turbulence is similar to the magnetic energy, because the internal energy is almost constant during the evolution.
We have found that when the initial $\epsilon_B$ immediately behind the shock is a few times $10^{-3}$, $\epsilon_B$ can grow on the order of 0.1 at maximum (see Fig. \ref{f6}).
\item The critical length scale below which the back reaction of magnetic field on turbulence becomes effective is $\sim 1/10$th the initial inhomogeneity scale.
At this scale, the magnetic energy becomes comparable to the kinetic energy, and the Kolmogorov cascade transitions to the MHD critical balance cascade.
In addition, below this scale, spatial correlations of the local magnetic field orientation appears (see \S\ref{Bstructure}).
\item For relativistic supersonic turbulence, the kinetic energy decay begins an order of magnitude faster than for the transonic case, because the formation of a number of shock waves directly dissipates the kinetic energy.
The evolution of $\epsilon_B$ is completely different from the transonic and subsonic cases--it decreases even while the magnetic energy is growing because the increase in the internal energy by shock dissipations is faster than the magnetic field amplification.
\end{itemize}

In the remainder of this section, we compare our results with previous related studies.
Using three-dimensional relativistic MHD simulations, \cite{ZMW} recently showed that the turbulence induced by the Kelvin-Helmholtz instability (KHI) in a relativistic shear flow amplifies the magnetic field.
They found that $\epsilon_B$ converges to $5\times 10^{-3}$ irrespective of its initial values of $10^{-5}$ and $10^{-7}$.
Our results, on the other hand, show that $\epsilon_B$ depends significantly on the initial value.
The difference between the two results stems from the sources of turbulence.
In the simulations of \cite{ZMW}, the shear flow is continuously injected by hand, which constantly drives the turbulence via the KHI.
On the other hand in our simulations, the turbulence that is expected to arise from the RMI is given only initially.
Considering these differences, we find that the results of the two simulations are essentially the same in the sense that the magnetic energy is deposited by turbulent flows induced by a hydrodynamic instability that saturates once it becomes as large as kinetic energy, and then the magnetic energy evolves along with the kinetic energy (see, Fig. 2 of Zhang et al. 2009 and Fig.~\ref{f5} herein).

The exponential growth of magnetic energy and the subsequent power-law decay obtained from our simulations is similar to the case of the Weibel instability \citep{CSA, KKSW}.
The Weibel instability is a powerful mechanism for the generation and amplification of magnetic fields.
However, the typical scale of the magnetic field on the order of the plasma skin depth may be insufficient to contribute to particle acceleration.
In addition, the magnetic fields generated at shock front decay rapidly.
The advantage of magnetic field amplification by the RMI is in its spatial and time scales.
Because the scale of the RMI is determined by that of the preshock density fluctuations, the typical scale of magnetic field can be macroscopic and typically 
comparable to the causally connected scale (i.e., the maximum scale in the observable region).

\section{DISCUSSION: IMPLICATIONS FOR GAMMA-RAY BURSTS}\label{sec:GRB}

The results of our simulations can address various issues.
For example, we consider here the implications for the GRB emissions.
The initial conditions of our simulations in Table 1 correspond to the GRB-emission region particularly to the internal shocks of GRBs with the following physical quantities:
\begin{itemize}
\item The kinetic luminosity is $L=4\pi r^2 m_{p} n c^3 \Gamma^2 \sim 6 \times 10^{52}$ erg s$^{-1}\ (r/10^{14}\ {\rm cm})^{2} (n/10^{10}\ {\rm cm}^{-3})$ $(\Gamma/10^3)^2$, which is a typical value, where $\Gamma$ is the bulk Lorentz factor.

\item The magnetic energy, carrying a fraction $\sim \epsilon_{B,0}$ of the total luminosity, is subdominant.
Considering the frozen-in magnetic field transported from the central engine $B_{\rm fz}$, the ratio of the magnetic to the total luminosity, $L_{B_{\rm fz}}/L = 4\pi r^2 (B_{\rm fz}^2/8\pi) c \Gamma^2/L= B_{\rm fz}^2/8\pi\,m_{p}\,n\,c^2 \equiv \epsilon_{B_0}$, is nearly conserved during the free expansion of the fireball because the area normal to the toroidal magnetic field is proportional \footnote{The area normal to the toroidal field evolves $\propto r^{-2}$ (not $\propto r^{-1}$) at $r \gtrsim  r_0 \Gamma^2 \sim 10^{13}$ cm because the shell thickness broadens. However, the shell broadening leads to the internal shocks between the successive shells, which prevent more than twice broadening.}  to $\propto r^{-1}$; hence, the comoving toroidal field evolves as $B_{\rm fz} \propto \Gamma^{-1} r^{-1}$.
The magnetic fraction $\epsilon_{B_0}$ corresponds to a central-engine magnetic field of
\begin{eqnarray}
B_0&\sim& \left(\frac{8\pi \epsilon_{B_0} L}{4\pi r_0^2 c}\right)^{1/2} \nonumber\\
&\sim& 3 \times 10^{14}\ {\rm G}\ 
\epsilon_{B_0}^{1/2} 
\left(\frac{L}{10^{53}\ {\rm erg}\ {\rm s}^{-1}}\right)^{1/2}
\left(\frac{r_0}{10^7\ {\rm cm}}\right)^{-1}. \label{eq:B0}
\end{eqnarray}
The preshock magnetic fields assumed in our simulations correspond to an initial fireball at $r=r_0$ being magnetized as $B_0 \sim 10^{12}$ G for run B1, $B_0 \sim 10^{11}$ G for runs B2, B2-s, and B2-r and $B_0 \sim 10^{10}$ G for run B3.
In these estimations, we have considered the effect of compression by internal shock ($\epsilon_{B, \rm ps}\simeq B_{\rm ps}^2/8\pi\,p_{\rm ps}\simeq 100\,\epsilon_{B_0}$, where the subscript ``ps'' indicates the values of the post internal shock or the initial condition of the series of run B).

\item The comoving temperature of the freely expanding fireball is adiabatically cooled to 
\begin{eqnarray}
k_{\rm B} T &\sim& k_{\rm B} T_0 \Gamma^{-1} (r/r_0 \Gamma)^{-2/3} \nonumber\\
&\sim& 2\,{\rm eV}\,\left(\frac{r}{10^{14}\ {\rm cm}}\right)^{-2/3}\left(\frac{L}{10^{53}\ {\rm erg}\ {\rm s}^{-1}}\right)^{1/4} \nonumber\\
&\times&\left(\frac{r_0}{10^7\ {\rm cm}}\right)^{1/6}\left(\frac{\Gamma}{10^3}\right)^{-1/3}, \label{eq:preT}
\end{eqnarray}
whereas it rises to $\sim 1\ {\rm GeV}$ after the internal shocks (followed by a similar adiabatic cooling).
In the simulation, we initially consider $k_B T \sim 9.38$ MeV to ensure the numerical stability.
However, this does not affect our conclusions as long as the initial temperature is well below the postshock temperature.

\item We expect a density inhomogeneity in a GRB jet because the angular size of a causally connected region $\Gamma^{-1} \sim 10^{-3}$ is usually smaller than the jet opening angle $\theta_j \sim 0.1$.
The density inhomogeneity is also suggested by the observations such as the large variation in the prompt luminosity compared to that in the afterglow \citep{KP00}, the spectral and temporal varieties \citep{IN01,YIN04}, and the variabilities of the early afterglow \citep{IKZ05} and its polarization \citep[][and references therein]{T09}.
The comoving size of the causally connected region is $\sim r/\Gamma \sim 10^{11}$ cm $(r/10^{14}\ {\rm cm}) (\Gamma/10^3)^{-1}$, which may be the typical scale of the density fluctuations.
The fluctuation scale might be smaller than this scale, because the sound velocity is less than the light speed $c$ before the shocks.
In either case, we may take $\lambda_c=L_z/3$ ($\sim$ simulation box size) as the fluctuation scale because the simulation is scale free.
\end{itemize}

In the following, we consider two leading models of GRB prompt emission: the synchrotron model and the photosphere model.
We also apply the afterglow model.

\subsection{Synchrotron Model}\label{sec:sync}
The internal shocks convert kinetic energy into internal energy, which goes into the magnetic field and the electron acceleration with energy fractions $\epsilon_B$ and $\epsilon_e$.
The electrons radiate synchrotron emission that is observed as the prompt GRB.
This is the internal-shock synchrotron model \citep{mes06}.

The characteristic synchrotron frequency is 
\begin{eqnarray}
\nu_m=\frac{\Gamma \hbar \gamma_m^2 eB}{m_e c}
&\sim& 2\ {\rm MeV}\
\left(\frac{\epsilon_B}{10^{-2}}\right)^{1/2} 
\left(\frac{f_e^{-1} {\cal R}^{-1} {\bar \Gamma} \epsilon_e}{10}\right)^{2}
\nonumber\\
&\times&
\left(\frac{L_{\gamma}/\epsilon_e}{10^{53}\ {\rm erg}\ {\rm s}^{-1}}\right)^{1/2}
\left(\frac{\Gamma}{10^{3}}\right)^{-2}
\left(\frac{\Delta t}{10^{-2}\ {\rm s}}\right)^{-1},
\label{eq:num}
\end{eqnarray}
where $f_e$ is the fraction of electrons that are accelerated, ${\cal R}$ is the number ratio of electrons (plus positrons) to protons, $\bar \Gamma$ is the relative Lorentz factor between shells, $\Delta t$ is the variability time, $\gamma_m \approx f_e^{-1} {\cal R}^{-1} {\bar \Gamma} \epsilon_e (m_p/m_e)$ is the characteristic random Lorentz factor of electrons, and we use $4 \pi r^2 (B^2/8\pi) c \Gamma^2 = \epsilon_B L_{\rm int} \approx \epsilon_B L_{\gamma}/\epsilon_e$ and $r=2 \Gamma^2 c \Delta t$.
In the synchrotron model, we identify the characteristic frequency with the observed peak energy of the Band spectrum \citep[e.g.,][]{ZM02}.
Our simulations indicate that the RMI can yield a sufficient magnetic fraction $\epsilon_B \sim 10^{-2}$ to reproduce the observed peak energy.
The necessary condition is that the initial magnetic fraction immediately behind the shock is $\epsilon_{B,{\rm ps}} \gtrsim 10^{-4}$ (see Fig.~\ref{f6}), i.e., the central engine magnetic field is $B_0 \gtrsim 10^{11}$ G from Eq.~(\ref{eq:B0}).
In other words, the Poynting flux may be subdominant.
It is interesting that the maximum level of magnetic energy depends on the initial magnetic field in the turbulence caused by the RMI instability (see \S~\ref{sec:decay}).
We emphasize that a strong requirement for the amplitude of the initial density fluctuations is not necessary because the hundredfold growth of magnetic energy can be realized irrespective of the velocity dispersion of turbulence (see, eq. [\ref{Bamp}]).
As the dispersion of the density fluctuations is reduced, the velocity dispersion decreases, which results in a longer timescale for amplification.
Because the local magnetic field strength is distributed as shown in Fig.~\ref{f8}, the typical magnetic fraction determines the peak energy.
Note that the factor $f_e^{-1} {\cal R}^{-1} {\bar \Gamma} \epsilon_e \sim 10$ in Eq.~(\ref{eq:num}) may require an efficient electron acceleration $\epsilon_e\sim 1$, a large relative Lorentz factor $\bar \Gamma \sim 10$, few positrons ${\cal R}\sim 1$, and/or a small fraction of accelerated electrons $f_e \sim 0.1$ \citep[e.g.,][]{EIC05,TIN08}.

Electrons should be scattered by disturbed magnetic fields, which is required for the first-order Fermi acceleration.
The relevant wavelength to resonantly scatter electrons with energy $\gamma_m m_e c^2$ corresponds to a Larmor radius of
\begin{eqnarray}
R_{m} \equiv \frac{\gamma_m m_e c^2}{e B} 
&=&7 \times 10^4 \ {\rm cm} \,
\left(\frac{\epsilon_B}{10^{-2}}\right)^{-1/2} 
\left( \frac{f_e^{-1} {\cal R}^{-1} {\bar \Gamma} \epsilon_e}{10} \right) \nonumber \\
&\times&\left(\frac{L_{\gamma}/\epsilon_e}{10^{53}\ {\rm erg}\ {\rm s}^{-1}}\right)^{-1/2}
\left(\frac{\Gamma}{10^{3}}\right)^{3}
\left(\frac{\Delta t}{10^{-2}\ {\rm s}}\right).
\end{eqnarray}
If this length-scale is attributed to the RMI instability, the initial size of the inhomogeneity is required to within a range $\sim 10$ to $1.0$ times $R_m$, which is much shorter than $r/\Gamma$.
However, even if the scale of the density inhomogeneities is much larger than $R_m$, the magnetic field fluctuations induced by the RMI can scatter the electrons via magnetic mirror reflections.
Indeed, \cite{BYL} recently studied the transport of test particles in MHD turbulence, and found effective scattering of particles by magnetic bottles formed by large-scale slow-mode perturbations.

The internal-shock synchrotron model has several crucial problems, one of which is the cooling problem \citep{MR00}.
The cooling time for electrons with the characteristic Lorentz factor $\gamma_m$ is usually much shorter than the comoving causal time of $\sim \Gamma \Delta t$, so that almost all electrons cool down to a Lorentz factor $\gamma_c \sim 6\pi m_e c/\sigma_T B^2 \Gamma \Delta t$.
The corresponding cooling frequency $\nu_c = \Gamma \hbar \gamma_c^2 eB/m_e c$ is
\begin{eqnarray}
\nu_c 
&\sim& 0.9\ {\rm keV}\
\left(\frac{\epsilon_B}{10^{-2}}\right)^{-3/2}
\left(\frac{\Gamma}{10^{3}}\right)^8
\nonumber\\
&\times&
\left(\frac{L_{\gamma}/\epsilon_e}{10^{53}\ {\rm erg}\ {\rm s}^{-1}}\right)^{-3/2}
\left(\frac{\Delta t}{10^{-2}\ {\rm s}}\right),
\label{eq:nuc}
\end{eqnarray}
below the characteristic frequency $\nu_m$.
In this case, the low-energy spectral index below the peak energy ($\nu_m$) becomes $F_{\nu} \propto \nu^{-1/2}$, which contradicts the harder observations $F_{\nu} \propto \nu^{0}$.

RMI turbulence could solve the cooling problem by continuously accelerating electrons through the stochastic acceleration so-called
the second-order Fermi acceleration in non-relativistic cases.
The quasi-linear theory for electron scattering gives us the scattering timescale as $t_{\rm sct}^{-1}=(\pi/4)f_{\rm R} \Omega_{\rm L}$,
where $f_{\rm R}$ is the energy density fraction of the resonant turbulence to the background magnetic field, and $\Omega_{\rm L}$ is the Larmor frequency.
The simulation results show the non-linear turbulences
($\langle |B| \rangle \sim \Delta |B|$),
and the non-resonant scattering may be essential as we mentioned before.
In spite of those issues, extrapolating this formula with $f_{\rm R} \sim 1$, the scattering timescale is estimated as $\sim \Omega_{\rm L}^{-1}$, which is significantly shorter than the cooling timescale.
Since the mildly relativistic turbulence leads to the energy variance per scattering $\Delta E/E \sim \Delta v/c$, the heating timescale due to the turbulences may be siginificant.
\cite{AT} demonstrated that the second-order Fermi acceleration balances synchrotron cooling, so that the low-energy spectral index becomes as hard as $F_{\nu} \propto \nu^{1/3}$--$\nu^0$, consistent with the observations.
In their Monte Carlo simulations of acceleration, the mean collision time of pitch-angle scattering is assumed to be independent of electron energy.
According to the measurement of the test-particle collision frequency in MHD turbulence by \cite{BYL}, the collision frequency is independent of energy because the particles are mainly scattered by magnetic bottles formed by large-scale slow-mode perturbations.
Thus we can expect the second-order Fermi acceleration in the post-shock turbulent region. 
\cite{AT} also demonstrated that if the electron collision frequency decreases with time, the accelerated electrons can reproduce the Band spectrum, including its high-energy side.
Their assumption is compatible with our simulation results, because the turbulence we are discussing is time dependent; its magnetic energy decays with time at a later stage, which leads to decreasing collision frequency.
The direct numerical simulations including the particle acceleration, which would be feasible by applying the test-particle approximation, is necessary to quantitatively confirm these expectations.

Another possibility to solve the cooling problem might be to consider that the eddy scale (i.e., the density fluctuation scale) is much smaller than the comoving causal scale.
As shown in Fig.~\ref{f5}, the magnetic field decays within a few eddy-turnover times.
If the decay timescale is comparable to the cooling timescale of electrons of $\gamma_m$, emissions from cooled electrons are suppressed.
Then, the low-energy spectral index becomes that of the synchrotron, $F_{\nu} \propto \nu^{1/3}$, consistent with the observations \citep{pee06}.
However note that the radiative efficiency becomes too small unless the decay times is fine tuned to the cooling time.

We also comment on the jitter radiation \citep{M00}, in which small-scale turbulences lead to average deflections much smaller than the beaming angle and a low-energy spectral index $F_{\nu} \propto \nu$ harder than that for the synchrotron, as observed in a fraction of GRBs.
Because the Larmor radius $R_L \sim \gamma_e m_e c^2/eB$, or more precisely $R_L/\gamma_e$, is much smaller than the typical magnetic field scale, which is roughly one-tens of the scale of initial density fluctuations (below which the power spectrum decreases as shown in Figs.~\ref{f7} and \ref{f10}), the jitter radiation does not work for macroscopic RMI turbulence.

An interesting prediction of the synchrotron model with RMI turbulence is the polarization of the prompt GRB.
Since the typical scale of the magnetic field is $\sim L_z/30$ (i.e., $\sim 1/10$th the fluctuation scale of $\sim L_z/3$), the number of domains with coherent magnetic field would be at least $N \sim 10^3$.
Therefore, the polarization degree should be less than
\begin{equation}\label{eq:polari}
\Pi_L \le \frac{70 \%}{\sqrt{N}} \sim 2 \%,
\end{equation}
\citep{GW99}, which may be probed by the future X-ray polarimetric observations \citep{T09}.
If the polarization of the prompt GRB is above this limit, we have to consider mechanism other than the synchrotron model via RMI turbulence,
such as the magnetic field advected from the central engine.

Even if the Band component at MeV range is produced by a mechanism other than synchrotron (e.g., photosphere emission, which is discussed in the next section), the extra high-energy component recently identified by the Fermi satellite \citep{LAT:2010us,Abdo:2009pg} may require emissions from non-thermal particles.
The extra components can be explained by several models such as early onset of the afterglow \citep{ghis09,kum09}, upscattering of external/photospheric photons \citep{TOM09b,TOM10,pee10}, and hadronic pair cascade \citep{agm09,asa10}.
For those models, the magnetic field amplification and turbulences to produce non-thermal particles are indispensable.
Another interesting method to emit GeV photons is the internal shock synchrotron with high Lorentz factor recently proposed by \citet{I10}.
In such cases the cooling frequency can reach $\nu_c \gtrsim 100$ GeV if the Lorentz factor is as high as $\Gamma \gtrsim 10^{4}$ in Eq.~(\ref{eq:nuc}).
With eq.~(\ref{eq:num}) ($\nu_m \propto \Gamma^{-2}$), the rising segment of the $\nu F_{\nu} \propto \nu^{(3-p)/2}$ spectrum is stretched down below $1$ keV over more than 7 energy digits, as observed.
In the high Lorentz factor model, the maximum synchrotron frequency is limited by the magnetic field decay time $f_B \Gamma \Delta t > \kappa \gamma_e m_e c/q_e B$, where $f_B$ is the ratio of the decay time to the comoving causal time and the right hand side is the electron acceleration time (with $\kappa \sim 1$ for the Bohm limit) \citep{I10}.
This yields
\begin{eqnarray}
\nu_{\max}^{\rm Bdecay} &\sim& 100\ {\rm GeV}\
f_B^{2} \kappa^{-2}
\left(\frac{\epsilon_B}{10^{-2}}\right)^{3/2}
\left(\frac{\Gamma}{10^4}\right)^{-6}
\nonumber\\
&\times&
\left(\frac{L_{\gamma}/\epsilon_e}{10^{53}\ {\rm erg}\ {\rm s}^{-1}}\right)^{3/2}
\left(\frac{\Delta t}{10^{-2}}\right)^{-3}.
\end{eqnarray}
Therefore, to produce high-energy photons, $f_B \sim 1$ is necessary, i.e., the eddy-turnover time $t_{\rm eddy} \sim L_z/\Delta v$ should be comparable with the causal time.

\subsection{Photosphere Model}

Another problem with the internal shock synchrotron model is an efficiency problem
\citep{kum99,Z07,ITYN06}.
The observed high-radiative efficiency requires large dispersion for the Lorentz factor
\citep{kob01}, which tends to destroy the observed correlations, $\nu_m \propto L_{\gamma}^{1/2}$ \citep{Y04}, because the peak energy $\nu_m$ is also sensitive to $\bar \Gamma$ and $\Gamma$ in Eq.~(\ref{eq:num})
\citep{asa03}.

The difficulties of the internal shock models lead to the re-examination of the original fireball model, in which photons are released in a photospheric emission when the fireball becomes optically thin \citep[e.g.,][]{MR00, RR05, TMR07,RP09,I07,I10}.
The original problem is alleviated by introducing dissipation under the photosphere, which can bring the thermal peak into the observed range.
The photosphere model can naturally achieve the high efficiency
and the hard low-energy spectrum. 
The only crucial flaw is that the spectrum tends to be thermal without the nonthermal tails observed in GRBs.
The electrons with mildly relativistic temperature can upscatter the thermal photons to produce a high energy tail (\citet{BEL10}, see also similar simulations for AGNs, \citet[][]{asa07,asa09}), but the mechanism to keep
electron temperature higher than photon temperature is not self-evident.

A probable solution is the RMI turbulence excited just below the photosphere, which could convert the thermal spectrum into the observed nonthermal spectrum through Comptonization.
This is similar to the \citet{T94} model, which employs Alfv\'en turbulence rather than the RMI turbulence.
If the turbulent velocity is mildly relativistic and the Thompson optical depth is about unity (i.e., just below the photosphere), the Compton $y$-parameter is approximately unity.
In this case, if the turbulent kinetic energy is comparable or larger than the radiation energy, photons are statistically upscattered above the peak energy into a broken power-law (Band-like) spectrum.
Note that electrons are, in a sense, continuously heated because they move together with protons in macroscopic turbulence, i.e., $\epsilon _e\approx 1$ is effectively achieved.
For the high-radiative efficiency of GRB, it is necessary that the turbulent energy is just comparable with the radiation energy.
In the RMI turbulence model, this is naturally achieved by the relativistic shock just below the photosphere, where the density is so low that the internal energy (i.e., microscopic motion of matter) is not effectively thermalized into radiation in the post shock.
Hence, the pressure balance between the radiation and the matter behind the shock automatically leads to near equipartition of radiation and turbulent energy \citep{I10}.

According to the simulation presented in Fig.~\ref{f2}, the turbulent velocity dispersion is larger for larger density fluctuations and its maximum is $\Delta v \sim 0.6 \times c/\sqrt{3} \sim 0.3 c$ (run A1).
In this case, the Compton $y$-parameter is
\begin{equation}
y=\tau_T \frac{4kT_e}{m_e c^2}
\sim 1 \left(\frac{\Delta v}{0.3 c}\right)^2
\left(\frac{\tau_T}{5}\right),
\end{equation}
where the dependence on $\tau_T$ is not square because the fireball is expanding with decreasing $\tau_T$.
The high-energy spectral index in $F_{\nu} \propto \nu^{1-\beta_B}$ is given by the unsaturated Compton spectrum \citep[e.g.,][]{RL04} with
\begin{equation}
1-\beta_B \sim 
\frac{3}{2}-\sqrt{\frac{9}{4}+\frac{4}{y}} \sim -1,
\end{equation}
as observed in the density fluctuation $\Delta n/n_0 \sim 1$ (i.e., $\Delta v \sim 0.3 c$) just below the photosphere $\tau_T \sim 5$.

Note that the photosphere model does not explain the high-energy photons above $\sim \Gamma m_e c^2 \sim 1\ {\rm GeV}\ (\Gamma/10^3)$.
However the high-energy emission can be produced by the subsequent emission such as the internal shock and afterglow emission (see Sec.~\ref{sec:sync}).

\subsection{Afterglow}\label{sec:afterglow}

The afterglow is thought to be produced by the relativistic shock between the outflow and the ambient medium via synchrotron emission, although the interpretation of early afterglows has not been settled \citep{Z07,ITYN06}.
The broadband modeling suggests various magnetic fractions of $10^{-5} \lesssim \epsilon_B \lesssim 10^{-1}$ with a mean of roughly $\epsilon_B \sim 10^{-2}$ \citep{PK02}.
The magnetic field amplification by the compression of $\sim \mu$G circumburst magnetic field merely yields $\epsilon_B \sim 10^{-9}$, which is too weak for the afterglow emission.
The small-scale field produced by the plasma instabilities such as the Weibel instability decays rapidly and does not persist over the emission region \citep{CSA,KKSW}.

The RMI turbulence dynamo could be responsible for the magnetic field generation of the afterglow.
In this mechanism, the maximum magnetic energy depends on the initial conditions (see \S\ref{sec:evolve}).
If the initial magnetic field is $\epsilon_B \sim 10^{-9}$ as expected for the compression of a circumburst magnetic field, the maximum $\epsilon_B$ would be $\ll 10^{-4}$ because the maximum would be smaller than than that for run B3 whose initial $\epsilon_B$ is $\sim 10^{-7}$.
Our maximum magnetic energy level is less than that expected in \cite{GM}, where almost all the kinetic energy induced by the density bump-shock interaction was supposed to be converted into the magnetic energy.
However, other mechanisms such as the cosmic-ray or secondary e$^\pm$ pair streaming instability \citep{LB,ram07} could preamplify the magnetic field to a moderate level $\epsilon_B \gtrsim 10^{-5}$, which can be boosted to the necessary level $\epsilon_B \sim 10^{-2}$ by the RMI turbulence dynamo.
The advantage of this scenario is that preamplification may be moderate and not necessarily complete.
In addition, the dependence of the initial conditions can diversify to the magnetic fraction $\epsilon_B$ as inferred from the observations.

RMI turbulence could also cause the shallow decay phase, which is the most enigmatic feature in the early afterglow.
The point is that the maximum value of $\epsilon_B$ depends on the initial density fluctuation if the the magnetic energy immediately behind the shock is more than two orders of magnitude smaller than the kinetic energy of turbulence (see, eqs.~[\ref{Bamp}] and [\ref{BK}]).
In addition, depending on $\Delta n/n$ and its initial scale $\lambda_0$, the magnetic field growth timescales ($\sim t_{\rm eddy} \sim \lambda_0/\Delta v
\sim \lambda_0 n/\Delta n/c_s$) can be longer than the dynamical timescale $R/(c \Gamma)$.
This effect may also lead to the effective $\Delta n/n$ dependence of $\epsilon_B$.
The stellar wind \citep[e.g.,][]{CMW75} or the ionizing radiation \citep[e.g.,][]{B89} from the GRB progenitors may diminish the nearby density fluctuations, or induce some hydrodynamical instabilities \citep{O88,ram05}.
Such radial dependencies of $\Delta n/n$ may effectively cause the evolution of $\epsilon_B$ (and probably also $\epsilon_e$) as discussed in \cite{Y03} and \cite{ITYN06} to explain the shallow decay phase of the early afterglow.

The observed afterglow polarization of $\Pi_L \sim 1\%$ \citep[e.g.][]{cov99,gre03}
is consistent with the theoretical upper limit of the RMI turbulence origin in Eq.~(\ref{eq:polari}).
If the polarization is actually caused by the RMI turbulence, it implies that the density fluctuation scale is comparable with the causal scale, which may be effectively realized by the inhomogeneous or structured outflow driving into the (even uniform) ambient medium.
A few events with $\Pi_L \sim 10\%$ such as GRB 020405 \citep{ber03} call for other mechanisms such as the magnetic field advected from the central engine.

The relativistic turbulence model was invoked to address the prompt emission by the afterglow shock \citep{nar09}.
Our simulation does not seem to support this picture because the turbulence begins to dissipate 
much faster than the eddy-turnover time (see \S\ref{sec:rela}).

\acknowledgments
Numerical computations were carried out on XT4 at the Center for Computational Astrophysics (CfCA) of National Astronomical Observatory of Japan.
This work is supported by a Grant-in-aid from the Ministry of Education, Culture, Sports, Science, and Technology (MEXT) of Japan, Nos. 22$\cdot$3369 (T. I.), and 22740117 (K. A.), and 19047004, 21684014, 22244019, and 22244030 (K. I.).

\clearpage

\end{document}